# Factors of people-centric security climate: conceptual model and exploratory study in Vietnam

**Duy Dang-Pham**
School of Business IT and Logistics
RMIT University
Melbourne, Australia
Email: duy.dang@rmit.edu.au

**Siddhi Pittayachawan**
School of Business IT and Logistics
RMIT University
Melbourne, Australia
Email: siddhi.pittayachawan@rmit.edu.au

**Vince Bruno**
School of Business IT and Logistics
RMIT University
Melbourne, Australia
Email: vince.bruno@rmit.edu.au

## Abstract

There is an increasing focus on the persuasive approach to develop a people-centric security climate where employees are aware of the priority of security and perform conscious security behaviour proactively. Employees can evaluate the priority of security as they observe and interact with the security features that constitute the security climate of the workplace. We examined the fundamental challenge that not every employee could recognise those features. In this multi-stage research, we adopted the theoretical lens of symbolic interactionism to advance a conceptual model which explains the relationship between organisation's social networks and the formation of information security climate. A descriptive case study in Vietnam was then conducted to refine the proposed model. The findings validated and extended the dimensions of information security climate, as well as identified the relevant organisation's social networks (i.e. information, affect, and power) that lead to its formation.

## 1 Introduction

Over the past decades, a large amount of theoretical and empirical researches have contributed findings about the users' important role in ensuring organisational information security. These studies investigated perceptions of the environmental cues that would result in security behaviours that are desired (e.g. Dang-Pham and Pittayachawan 2015; Herath and Rao 2009; Siponen et al. 2014) or undesired by the organisations (e.g. D'Arcy and Devaraj 2012; Dang 2014; Guo and Yuan 2012). Even though important factors that motivate security practices have been determined, there lacks investigations of how to exploit these factors to develop and maintain an information security environment. To answer this question, we investigate how information security perceptions could be disseminated amongst the employees, which subsequently form an information security workplace.

From the industry perspective, the development of people-centric security workplaces was mentioned as an emerging trend (Gartner 2015). People-centric security workplaces focus on building a group culture that fosters acceptable security behaviours, as well as the individual's perceptions of accountability and responsibility, by addressing their interrelated actions in the workplace's social networks (Gartner 2015). Academics have also examined security interactions amongst individuals such as delegation of responsibility and co-participation in performing security behaviours (Albrechtsen and Hovden 2010; Dourish et al. 2004; Thomson et al. 2006). Most recently, Dang-Pham et al. (2014) extended these arguments by highlighting the interactions among human and non-human entities (e.g. policy). These interactions describe how the employees could have access to and utilise the resources in their workplace for making security decisions. Furthermore, Dang-Pham et al.

(2014) suggested future studies to employ social network analysis techniques to investigate those interactions and their contributing effects towards information security behaviours.

As we are interested in studying the development of a people-centric security workplace, this leads to the investigation of a relevant construct named information security climate. Information security climate has been conceptualised inconsistently by prior researches (Dang-Pham et al. 2015). These researchers also argued that there has been little effort to link information security climate with the workplace's social networks. As a result, the goals and implications of this study are twofold. First, we discuss the interactionist perspective in the context of information security climate and forward a conceptual model with hypotheses that can be tested with social network analysis techniques. Second, we present key findings from an exploratory case study that identified the interactions that disseminate information security perceptions in ISO 27001-certified companies in Vietnam. This exploratory study provides empirical evidences to refine our proposed conceptual model. Additionally, the case study's findings contribute to the current body of knowledge in which information security researches in non-Western contexts are lacking (Crossler et al. 2013). In summary, this is a multi-stage research, and it aims to answer the following questions:

**RQ1:** What is the relationship between organisational networks and perceptions of information security climate; and

**RQ2:** What are the organisational networks that influence perceptions of information security climate, as suggested by information security professionals in Vietnam?

## 2 Literature review

### 2.1 Conceptualising information security climate (ISC)

The concept of information security climate (ISC) originated from one of the earliest studies by Chan et al. (2005), who conceptualised ISC perception as reflecting the perceived organisational concerns about information security. Subsequent ISC researches continued to use parts of Chan et al.'s (2005) model and test ISC models differently (Goo et al. 2014; Jaafar and Ajis 2013). Dang-Pham et al. (2015) suggested that ISC construct may have four dimensions, namely (1) leadership, (2) job & role, (3) workgroup, and (4) organisational systems. Their proposed dimensions were influenced by the seminal works of James and colleagues (James et al. 2008; Jones and James 1979), that defined organisational climate as the shared employees' cognitive valuations of the workplace's events that are significantly meaningful to them. These dimensions are related to the personal values system, which James et al. (2008) argue that their factors (e.g. role ambiguity of "job & role" dimension) match with a person's desires (e.g. desire for clarity) and impact their workplace well being. As a result, employees would be more aware of these factors in the workplace than the others and make sense of them, and organisational climate could be measured by aggregating individuals' climate perceptions (James et al. 2008).

Another important area that could contribute to the conceptualisation of ISC is safety climate which shares many common features (Chan et al. 2005). Both of these concepts refer to the climates that focus on a specific facet of an organisation (i.e. information security and safety) (Schneider et al. 2013). Moreover, information security and safety are strategic goals that require the employees' compliance yet often conflict with competing work demands (Chan et al. 2005). Similar to James et al.'s (2008) discussion, safety climate was defined as the shared perceptions of the safety's policies, practices, and procedures that are rewarded and supported (Zohar 2013). While there is no consensus regarding safety climate's dimensionality, supervisor and co-workers' safety practices were found to appear repeatedly in safety climate questionnaires and could inform the priority of safety practices (Zohar 2013). Similarly, prior ISC researches (Chan et al. 2005; Dang-Pham et al. 2015; Goo et al. 2014; Jaafar and Ajis 2013) measured direct supervisor and co-workers' practices in their models even though the factorial structures were different. Given this similarity, we support conceptualising ISC as a higher-order, reflective construct that consists of these two dimensions. We also define ISC as the perceptions of information security policy, practices, and procedures that are supported and expected in the workplace. Ultimately, ISC informs the priority of information security in the workplace.

### 2.2 The formation of information security climate

The formation of organisational climates can be analysed by adopting the theoretical lens of the interactionist perspective (Schneider and Reichers 1983). This perspective draws on symbolic interactionism to explain how organisational climates could be socially constructed as the employees interact with their workgroup (Ashforth 1985). This explanation fits our conceptualisation of ISC which reflects the observed practices of co-workers and supervisors. Ashforth (1985) also discussed the important role of informational and normative influences in the formation of organisational climates. He explained that informational influence provides the information to reduce the ambiguity and uncertainty in the workplace, especially via social comparison and conformity mechanisms. Normative social influence offers prescribed behaviours and beliefs that regulate climate by removing disagreements.

These social influences provide a theoretical ground for identifying the relevant interactions that affect security beliefs and behaviours. We reviewed literatures about social influences and sense making in workplace, especially the studies that employed social network analysis techniques. Social network analysis techniques are particularly useful in this context by providing the criteria to determine the most relevant interactions amongst a plethora of potential ones (Dang-Pham et al. 2014, 2015). Interactions or networks can be determined based on their natures and transactional contents. On the one hand, researches have been examining the effects of networks whose natures are expressive (e.g. friendship) and instrumental (e.g. seek/give advice) on individual's perceptions (Sykes et al. 2009). On the other hand, Tichy et al. (1979) suggested four types of exchanged contents, including affect (e.g. liking), influence or power, information, and goods or services.

The interactionist perspective sees organisational climates as formed by informational and normative social influences that reduce ambiguity and uncertainty in the workplace. By defining ambiguity as a situation when an individual is confused by too many interpretations and uncertainty refers to the absence of information, Saint-Charles and Mongeau (2009) found that employees tend to seek experts (i.e. instrumental network) for reducing uncertainty while clarifying ambiguity with friends (i.e. expressive network). When information security policies are announced, employees would face both ambiguity (e.g. trying to interpret the priority of information security) and uncertainty (e.g. being required technical knowledge to perform secure practices) (Dang-Pham et al. 2015). In addition, the concept of power and authority has been studied extensively in influence researches, especially when compliance is coerced and voluntary choices may not be granted (Cialdini and Goldstein 2004). As a result, it is sensible to test instrumental (i.e. advice and support), expressive (i.e. friendship and affiliation), and power (i.e. influence) networks for their effects on the formation of ISC.

## 2.3 The mediators of information security climate and compliant intention

To increase our model's practical values, we added intention to comply with security policies as the outcome of the networks' features and ISC. These effects have been supported with empirical results of prior researches. For example, ISC dimensions (e.g. supervisor and peers' behaviour) were found to increase compliant intention (Goo et al. 2014) and actual compliance (Jaafar and Ajis 2013). Likewise, Sykes et al. (2009) extended Technology Acceptance Model and found networks' indices influence intention to adopt new information system. In addition, we argue that there would be variables in between that mediate those direct relationships.

First, data about the different types of social networks only informs the ties possessed by individuals that provide the resources for achieving certain goals. These ties serve as the facilitating conditions that affect individual's perceptions, such as the more knowledge sources one has, the more confident they would feel in coping with difficult tasks and adopting new technology (Sykes et al. 2009). In this example, such confidence could mediate the number of possessed ties and adoption intention. Second, ISC reflects the employees' perceptions of the expected security practices and policies, which in turn infer the priority of information security in the workplace. This perceived priority and its effect on performance have been confirmed in safety climate researches. For instance, the safety climate models of Neal et al. (2000) and Zohar (2013) describe "safety motivation" as reflecting safety's priority while being the mediator of safety climate and behaviour. Regarding ISC, only the model of Chan et al. (2005) reflects those mediated relationships. It is also worth noticing that both Jaafar and Ajis (2013) and Goo et al. (2014) found weak (b = 0.121) and unsupported hypotheses when testing the direct impact of ISC on compliant intention and actual compliance. In contrast, the effect of Chan et al.'s (2005) "ISC perceptions" construct as a mediator on compliant intention was stronger (b = 0.237). Such difference hints that adding mediators between ISC and the outcome variables could be a better solution.

We found the factors of Theory of Planned Behaviour (TPB) to be the most relevant for serving as the mediators in our conceptual model. TPB hypothesised that behavioural intention is motivated by attitude, subjective norm, and perceived behavioural control (Ajzen 1985). While TPB is widely known in social science researches, it is important to discuss the current arguments about the construct "Perceived behavioural control". Even though Ajzen (2011, p. 446) noted that perceived behavioural control is "very similar to Bandura's conception of self-efficacy", empirical evidences consistently showed that this factor would be more accurately conceptualised as a higher-order construct that reflects self-efficacy and perceived controllability (Hagger and Chatzisarantis 2005; Trafimow et al. 2002). We follow this suggestion as it adds richer explanations to the model. In particular, we argue that the perceived priority of information security (inferred by ISC and social interactions) would be described by the attitude about security's importance, as well as the belief that performing security practices is their own responsibility and control.

## 3  Conceptual model development

### 3.1  TPB and ISC hypotheses

TPB hypothesises that behavioural intention is motivated by three factors, including attitude, subjective norm, and perceived behavioural control (Ajzen 2011). This leads to the following hypotheses:

**H1:** Information security attitude positively affects Compliant intention

**H2:** Subjective norm positively affects Compliant intention

**H3:** Perceived behavioural control positively affects Compliant intention

Positive attitude about the priority of information security could be affected by ISC, as employees observe the espoused values being consistently enacted by their immediate colleagues and supervisors. Zohar (2013) explained based on expectancy theory's tenet that when safety is interpreted as a priority in the workplace, it invokes positive attitudes that make the employees adjust their behaviour while expecting rewards from doing so. Likewise, Neal et al. (2000) cited performance theory which posits that attitude about safety's priority mediates the relationship between organisational climate and safety performance.

Perceived behavioural control describes the confidence in completing tasks (i.e. self-efficacy) and the belief in having control over the target behaviour (i.e. perceived controllability). While prior ISC researches have not investigated the relationship between climate and self-efficacy, there were empirical results supported the link between safety climate and knowledge (Neal et al. 2000; Zohar 2013). Zohar (2013) reasoned that safety knowledge and skills become more desirable due to the prioritisation of safety climate and urge the employees to invest more effort in acquiring those skills. It is then reasonable to propose that employees who spent effort in enhancing information security knowledge and skills (by perceiving ISC) would feel confident in performing security behaviour.

Moreover, employees working in an environment where co-workers and supervisors actively follow the policy and take ownership of information security would be pressured by the norms and refrain from delegating responsibility to the others. The continuous observation of ISC would make one realise their ownership over information security tasks, as well as the subjective norm in relation to performing security practices. This argument has been confirmed by prior studies such as Liang et al. (2010) who found significant effect of team climate on subjective norm. As a consequence, the following hypotheses are proposed:

**H4:** ISC positively affects Information security attitude

**H5:** ISC positively affects Subjective norm

**H6:** ISC positively affects Perceived behavioural control

### 3.2  Network hypotheses

Our last set of hypotheses concerns the effects of social networks' features towards on ISC and TPB's factors. The formation of organisational climates and ISC was argued to result from the social interactions that facilitate sense-making (Ashforth 1985; Dang-Pham et al. 2015). Zohar and Tenne-Gazit (2008) found densities of communication and friendship networks significantly enhanced climate strength, which in turn supported symbolic interaction as an antecedent of safety climate. Possession of ties could also motivate individuals by providing resources and influences. As previously argued, one would perceive higher self-efficacy and control over security behaviour as they gain more knowledge about it. Individuals who have access to knowledge sources in the organisational networks would have more learning opportunities and perceive higher self-efficacy. That individual's possession of ties to knowledge sources can be measured as network centrality, which Dong and Yang (2009) demonstrated that structural capital centrality improved self-efficacy.

Attitude and realisation of norm could also be affected by informational and normative influences that are both subject to the flows of information and influence exchanged amongst network actors. Hossain and de Silva (2009) found individual's possession of ties affect attitude about technology adoption. Similarly, it was confirmed (Chow and Chan 2008) and theorised (ten Kate et al. 2010) that individuals having access to more extensive networks tend to have enhanced perception of subjective norm. Network theory also postulates that two individuals sharing similar beliefs and perceptions are due to diffusion (i.e. interacting with direct contacts) or structural equivalence (i.e. being connected with the same third person but not necessarily with each other) (Borgatti et al. 2013). Based on these arguments, it is logical to hypothesise that centrality and equivalence indices positively impact security attitude and subjective norm. The network hypotheses are proposed below and illustrated with the other hypotheses in Figure 1:

**H7:** Centrality & equivalence indices positively affect Information security attitude

**H8:** Centrality & equivalence indices positively affect Subjective norm

**H9:** Centrality & equivalence indices positively affect Perceived behavioural control

**H10:** Centrality & equivalence indices positively affect ISC

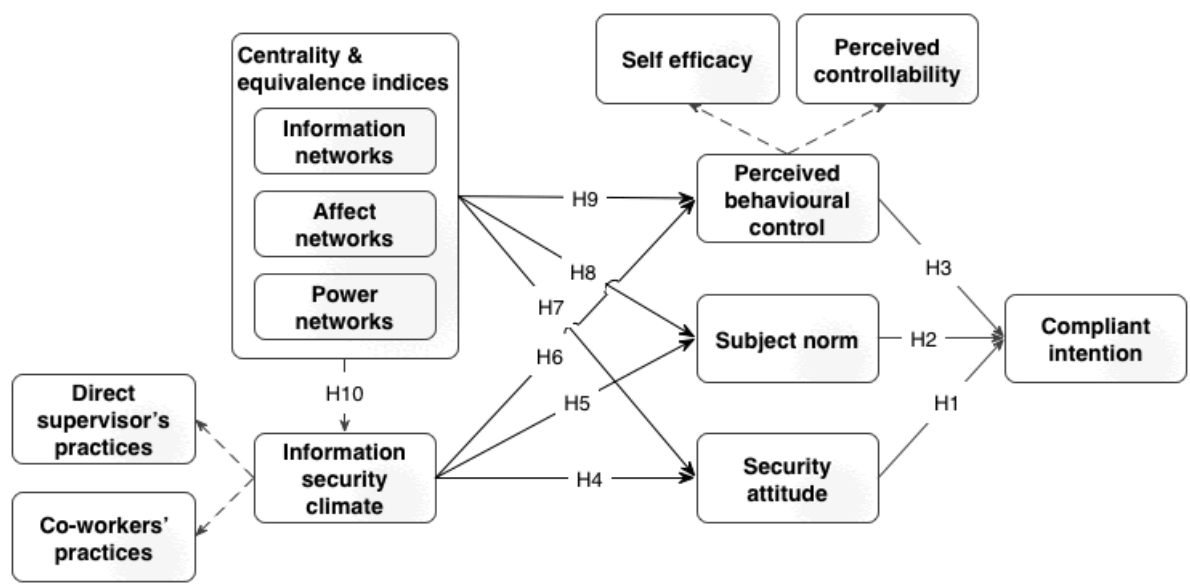

*Figure 1: Proposed conceptual model and constructs' dimensionality*

# 4　An exploratory study in Vietnam

## 4.1　Rationale & context for the descriptive case study

The conceptual model in Figure 1 specifies the constructs' dimensions whose measurements can be found from prior researches and used to develop instruments for data collection. The hypotheses can be tested quantitatively with regression analysis techniques. However, the social networks were added to the model based on theoretical suggestions as per our literature review. In addition, the adoption of social network analysis is uncommon in information security behavioural area which makes it challenging to select the meaningful ties from a plethora of organisational interactions (Dang-Pham et al. 2014). Dang-Pham et al. (2015) suggested that it is foremost necessary for network studies to identify the ties that are relevant to information security context. To refine our model, we conducted an exploratory, descriptive case study in Vietnam. The objectives of the case study include (1) identifying the relevant network ties and (2) validating the dimensions of ISC.

According to Yin (2003, p. 13), a case study is "an empirical enquiry that investigates a contemporary phenomenon within its real-life context, especially when the boundaries between phenomenon and context are not clearly evident". A descriptive case study specifically aims to present objectively the research phenomena rather than interpreting causations (Dubé and Paré 2003; Gerring 2004). Moreover, descriptive case studies are most effective for investigating areas that are under-researched (Laws and McLeod 2006). The concepts of social networks in people-centric security workplaces are complex and relatively new (Dang-Pham et al. 2014; Gartner 2015), thus case studies are necessary for keeping academic researches in pace with industry practices (Dubé and Paré 2003). Moreover, there is an emerging need for security research in non-Western contexts (Crossler et al. 2013). Provided our exploratory goal, the descriptive case study conducted in Vietnam as a South East Asian country is justified.

*Context:* since Vietnam is rapidly transforming into a global IT sourcing hub, local companies are pressured by business partners to enhance their information security management systems (ISMS). The Vietnamese ICT white book in 2014 reported that only 30 per cent of the local companies have implemented security policies and controls. In addition, 39 firms had their ISMSs certified for the international security standard ISO 27001 in 2013 (ISO 2015). For our research, we interviewed six security professionals who were project managers of six different ISO 27001's implementation projects (Table 1). All participants were working in companies in Ho Chi Minh City which is one of the most urban cities in Vietnam. This sample selection's strategy is to ensure that the interviewees have the expertise and knowledge to provide informed insights.

|    | Occupation | InfoSec experience | Industry |
|----|------------|--------------------|----------|
| P1 | InfoSec Consultant/Auditor | 3 years | Banking |
| P2 | IT Manager | 14 years | IT services |
| P3 | InfoSec Consultant | 5 years | Banking |
| P4 | Information Security Officer | 7 years | IT services |
| P5 | Deputy IT Director | 10 years | Banking |
| P6 | Data Security Manager | 3.5 years | Engineering and electronics |

*Table 1: Interviewees' profiles*

## 4.2 Addressing case study's research rigour

*Design:* Dubé and Paré (2003) recommended case study researchers to ensure research rigour by being transparent in design, data collection and analysis, and reporting results. Our design follows Gerring's (2004) structure of a descriptive case study which consists of case(s) that belong to a research unit. Under this structure, the network ties and dimensions of ISC in the people-centric security workplaces in Vietnam constitute a research unit. This unit comprises of interviews with six professionals that serve as cases.

*Data collection:* we initiated data collection process by sending interview invitations to the researchers' known contacts, online forums, and social media platforms such as LinkedIn for one month. We received seven acceptances to participate in the interview in the beginning, but one professional pulled out from the project after consulting with their management. As a result, six semi-structured interviews were conducted in-person in Vietnamese and audio-recorded, which each lasts one hour on average.

*Data analysis:* interviewed data was analysed by using NVivo 10 to perform thematic coding with social interactions, co-workers and supervisor's practices as major themes. Matching patterns emerged as we performed within- and cross-cases analyses (Yin 2009), which reinforce internal validity (Dubé and Paré 2003). External validity was achieved by having the participants reviewed and provided feedback about the analysed results, and we reported direct quotes in our coming discussions to ensure rigour (Dubé and Paré 2003; Riege 2003).

*Protocol:* presenting the case study protocol further enhances research rigour (Dubé and Paré 2003). Our protocol begins with performing a literature review to develop a descriptive theory that guides the case study and support its findings (Riege 2003; Zaidah 2007). Our descriptive theory is the first part of the conceptual model depicted in Figure 1, which consists of the network ties that were hypothesised to affect ISC and its dimensions. The case study protocol summarises our research process as illustrated in Figure 2 below.

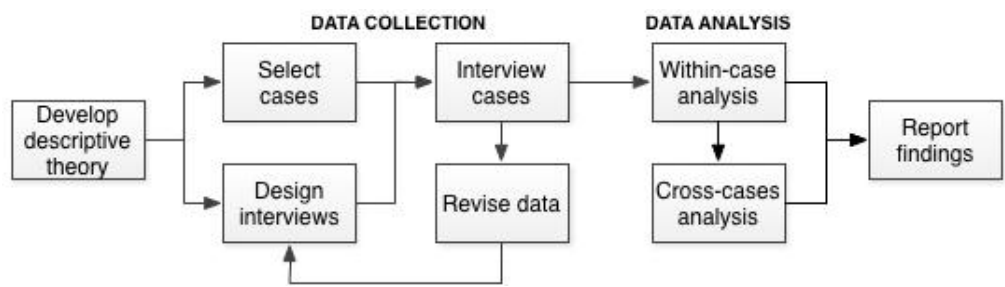

*Figure 2: Case study protocol*

# 5   Findings and discussion

## 5.1   Informational networks

Our interviews asked about the environmental features that promote the priority of information security and autonomous compliance, as reflecting a people-centric security workplace. The priority of security can be explained by having the employees attend security training or read security guidelines. They may also have the impression that security is prioritised from observing the workplace's ISC and perceiving the social pressures from co-workers and supervisor's practices. In other words, the ISC features emit informational and normative influences or networks that give rise to the priority of security.

Information about the importance of organisational security can be exchanged via the employees' interactions with security training programs and communications. In particular, the interviewed participants suggested conveying information about the relevant security threats and benefits of compliance to educate the users on the priority of security. However, explaining the benefits of compliance was considered a tricky task. Some participants argued that only the organisation could recognise the benefits of its employees' compliance. In contrast, employees could see the immediate consequences (e.g. data loss, sanction) more easily than at organisational level (e.g. reputation & financial losses). A recommended solution was to explain the benefits of compliance as mutual:

> "When organisations tell the users that there'll be security threats that cause reputation and financial losses if they don't comply, they would think: 'Who cares, in the worst scenario I'll just quit the company.'" (P2)

> "Your regular trainings must emphasise that if the company can demonstrate excellent information security through consistent compliance, we will gain trust from the clients and they would give us more large projects… and there could be pay raise from that too." (P4)

Alternatively, management can employ security communications such as posters, flyers, screen savers, announcements and reminder emails to promote the priority of security. These communications often include information about security threats that are relevant and realistic. While these communications are not guaranteed to change the users' security behaviours, their continuous deployment is cost-effective and may be effective in long term by overloading the users with information.

Besides learning about security's importance from security training and communications, the participants discussed that it would be most effective to have the employees interact and exchange information with their direct supervisors. Unlike generic security training and communications, direct supervisors know best the daily operations of their work units and can educate their subordinates on how to align security tasks with other work priorities. Moreover, they can persuade the subordinates more easily due to their closer ties in the workplace. The role of direct supervisors in promoting the priority of security is especially vital in large and hierarchical organisations:

> "If the firm is too large then the direct supervisors are responsible for hosting security talks with their subordinates. For example, they can share with their work unit the security issues that have occurred every month, and explain how those issues can affect the unit's KPI and everyone's job." (P3)

> "The end users would firstly observe how their department heads behave and think; if those leaders are well aware of the priority of information security then they will perform the secure behaviour, and in that case their subordinates wouldn't want to do differently." (P6)

However, the participants also warned about the issue of autonomous departments that allow security compromises or workarounds. As a result, they discussed necessary traits that direct supervisors should possess to effectively promote the priority of security:

> "To be a security leader, the direct supervisor should have leadership, communication skills, and the ability to lead by example besides having sound security expertise. They should be able to realise the security risks and understand their impacts on the business." (P2)

> "Direct supervisors need to understand clearly the security expectations of the top management. They also need to be able to motivate and inspire their subordinates to perform security behaviours, either by helping them to understand security in a friendly manner or simply by contiuously demonstrating their compliance." (P4)

### 5.2 Normative networks

Normative networks pressure the employees to realise the priority of security without necessarily understanding the benefits of compliance. Audit and monitoring, as well as security policies and sanctions were mentioned as the workplace's features that produce such pressures. While the participants acknowledged the importance of announcing the levels of sanctions in security policies, they did not suggest using sanctions to motivate compliance. The participants explained that sanctions could be perceived by the employees, especially the white-collar office workers, as a threat and would react against it. Moreover, participant P3 described sanctions as a reactive approach to manage organisational security, since effective ISMS should prevent the violations from happening rather than punishing them. The participants recommended conducting periodic and transparent audits and monitoring instead of using sanctions to deter violations and encourage compliance. Performing regular audits was described as enabling the organisation to recognise compliant efforts and detect potential violations to issue warnings. More important, these auditing efforts help to continuously remind the employees about their responsibility in ensuring organisational security and that they are accountable for such task.

Another normative source with which the employees can interact is their co-workers in the same work units. The participants discussed that the employees could interpret the priority of security in the workplace by observing their colleagues' behaviours such as voluntarily taking security considerations while performing daily work or displaying concerns about being held accountable for security consequences. In addition, the effects of socialisation and social exclusion could be observed from the interviews:

> "Even a person with high security awareness would eventually change when they join a firm where no one cares about security–all being carefree and having no rules to follow… For example, the co-workers would point at the person and ask: 'why would you spend your valuable time worrying about these security things?'" (P1)

Although the participants confirmed the high collectivism of Vietnamese culture in the workplace, they discussed that exploiting such cultural trait to disseminate security perceptions without purposeful interventions may be ineffective. For instance, E3 suggested that the effects of rewards and sanctions would be recognised easily by the employees in Vietnamese workplaces thanks to the widespread word-of-mouth network. However, the other participants did not find interpersonal influences caused by co-workers desirable:

> "Social influence is about persuading someone to like something and do it. The employees may like a person and follow him or her, but security is also about being enforced to follow the policies. Only the managers with formal authorities have the power to enforce but not anyone." (P2)

> "They could observe their colleagues perform security behaviours, and even though they know the good intention behind that, they won't necessarily follow because they don't feel like doing those behaviours… unless they see some benefits in doing so, but I don't think they would see any since performing security behaviours is not a competition after all." (P6)

## 6 Discussion & future directions

### 6.1 Organisational networks and information security climate

The exploratory findings validated and refined the proposed conceptual model (Figure 1). Our interviews found both informational and normative networks play important roles in promoting the priority of security. Informational networks describe information exchange between the employees and their direct supervisor and co-workers, as well as training sessions and security communications. In addition, periodic audit & monitoring, security policies, and co-workers were recognised as normative sources of the pressure that makes the employees believe in the priority of security. Affect (i.e. friendship) and power networks can be found in the exchange of such pressure between those normative sources and each individual. Furthermore, the findings suggested that non-human entities in the workplace (e.g. training session, policies, audit & monitoring) can be included in social network analysis as network actors besides co-workers and supervisor.

### 6.2   Dimensions of information security climate

The construct of ISC in Figure 1 was conceptualised to reflect observed co-workers and direct supervisor's security practices, based on prior literatures about ISC (Chan et al. 2005; Dang-Pham et al. 2015; Goo et al. 2014; Jaafar and Ajis 2013) and safety climate (Zohar 2013). The case study's findings confirmed and also extended this conceptualisation. Specifically, ISC components may also include training sessions, audit & monitoring, security policies, and security communications. These components fit the definition of ISC which covers the observable workplace's characteristics that promote the priority of security. Nevertheless, including them as dimensions of ISC would result in broader coverage of the construct at the expense of increasing the research instruments' length. In other words, more questions about ISC would need to be added to quantitative survey or qualitative interviews, which burden the participants. As a result, future studies that come up with a validated instrument for measuring ISC would be desired.

### 6.3   Implementation tactics to develop a people-centric security climate

People-centric security workplace consists of employees who are aware of the security's priority and their responsibility in performing security tasks (Gartner 2015). Such perceptions are represented by the construct ISC, and the interactionist perspective explains that descriptive meanings such as ISC emerge as employees interact with the information sources and interpret them (Ashforth 1985; James et al. 2008). The case study demonstrates both sense-making and sense-giving processes that give rise to ISC in the real context. Organisations then employ sense-giving tactics which exploit rituals and mentors or advisors to communicate the desired messages (Ashforth 1985). Security training programs and periodic audits, as well as the use of formal leaders that display the priority of security via leading by example, are illustrations of such tactics. As a consequence, the case study's findings could serve as practical recommendations for organisations to develop their people-centric security workplaces.

### 6.4   Compliance meaningfulness as an interpretation of workplace's security

The construct ISC in our research only reflects the descriptive meanings of co-workers and supervisor's practices that reflect the priority of security. Besides such descriptive meanings, psychological climate literatures suggested there exist subjective meanings that vary amongst individuals (James et al. 2008). One type of those subjective meanings is "work meaningfulness", which refers to the significance attached to one's work that is evaluated as important and meaningful (Rosso et al. 2010). According to these researchers, work meaningfulness has been found to affect important individual's outcomes such as commitment, engagement, behaviour, and performance. This construct could be conceptualised in security context as the meaningfulness of compliance, or how the employees perceive their compliance could play an important role and contribute to a more secure workplace. On the one hand, compliance meaningfulness is different from the descriptive meanings of ISC in that it reflects the characteristics of security tasks as perceived subjectively by individuals. On the other hand, they are relevant to each other and to the definition of people-centric security workplace as both constructs describe the employees' perceptions of the expected security practices and priority of security.

The participants in our case study emphasised on explaining the mutual benefits and consequences of compliance and violations to justify the meaningfulness of security. Investigating the meaningfulness of security is also aligned with the recent transition of security behavioural researches towards the persuasive approach to motivate compliance. For example, Karjalainen and Siponen (2011) proposed the experiential training approach that educates users on the need for security protection and convinces their voluntary compliance. We believe investigating compliance meaningfulness along with ISC could contribute to the whole picture of meanings in information security context and the persuasive approach in particular.

## 7   Conclusion

It can be observed that people-centric security workplace was most recently discussed as an industry trend (Gartner 2015) and security behavioural researches are moving towards the persuasive approach (Karjalainen and Siponen 2011). These two directions are aligned with each other by both aiming to develop an organisational environment where employees understand the priority of security and perform security behaviours proactively.

We examined the fundamental challenge that not every employee can have access to and recognise the security cues that are implemented by the organisation. Moreover, a workplace consists of security features whose meanings can be interpreted differently by the employees, and so is the priority of security. We adopted the theoretical lens of symbolic interactionism to investigate the formation of information security climate, which is defined as the descriptive meanings of the security features. Information security climate reflects the perceptions of security policies and practices that subsequently infer the priority of security.

In this multi-stage research, we forwarded a conceptual model which describes the formation of information security climate as a result of the employees' interactions with the organisation's social networks. The model's hypotheses were grounded on established theories and empirical findings of prior studies, and they can be tested with regression techniques and social network analysis. In the second stage, we conducted a descriptive case study with six information security professionals in Vietnam to refine the constructs of our proposed model. The limitations of this second stage include the small sample size of the descriptive case study as well as its scope which limited to Vietnam.

The exploratory findings confirmed co-workers and supervisor's practices as important dimensions of information security climate. Furthermore, they provided additional dimensions that include security training, policies, audit & monitoring, and security communications. These findings also suggested recommendations for developing people-centric security workplaces, especially in contexts that are similar to Vietnam as a developing country in South East Asia. More important, informational and normative networks were identified from the case study, which supported their inclusion in our conceptual model. Finally, by considering the effects of social networks in the model, we suggested the adoption of social network analysis as a novel technique to investigate security behaviours.

## References

Ajzen, I. 1985. *From intentions to actions: A theory of planned behavior*, Springer Berlin Heidelberg.

Ajzen, I. 2011. "Theory of planned behavior," in *Handbook of Theories of Social Psychology: Volume One*, p. 438.

Albrechtsen, E., and Hovden, J. 2010. "Improving information security awareness and behaviour through dialogue, participation and collective reflection. An intervention study," *Computers & Security* (29:4), Elsevier Ltd, pp. 432–445.

Ashforth, B. 1985. "Climate formation: Issues and extensions," *Academy of management review* (10:4), pp. 837–847.

Borgatti, S. P., Everett, M. G., and Johnson, J. C. 2013. *Analyzing Social Networks*, Sage Publications Ltd.

Chan, M., Woon, I., and Kankanhalli, A. 2005. "Perceptions of information security at the workplace: linking information security climate to compliant behavior," in *Perceptions of Information Privacy and Security* (Vol. 1), pp. 18–41.

Chow, W. S., and Chan, L. S. 2008. "Social network, social trust and shared goals in organizational knowledge sharing," *Information and Management* (45:7), pp. 458–465.

Cialdini, R. B., and Goldstein, N. J. 2004. "Social influence: compliance and conformity.," *Annual review of psychology* (55:1974), pp. 591–621.

Crossler, R. E., Johnston, A. C., Lowry, P. B., Hu, Q., Warkentin, M., and Baskerville, R. 2013. "Future directions for behavioral information security research," *Computers & Security* (32), Elsevier Ltd, pp. 90–101.

D'Arcy, J., and Devaraj, S. 2012. "Employee Misuse of Information Technology Resources: Testing a Contemporary Deterrence Model," *Decision Sciences* (43:6), pp. 1091–1124.

Dang, D. P. T. 2014. "Predicting insider's malicious security behaviours: a General Strain Theory-based conceptual model," in *2014 International Conference on Information Resources Management (Conf-IRM 2014)*, Ho Chi Minh City, Vietnam.

Dang-Pham, D., and Pittayachawan, S. 2015. "Comparing intention to avoid malware across contexts in a BYOD-enabled Australian university: A Protection Motivation Theory approach," *Computers & Security* (48), Elsevier Ltd, pp. 281–297.

Dang-Pham, D., Pittayachawan, S., and Bruno, V. 2014. "Towards a complete understanding of information security misbehaviours: a proposal for future research with social network approach," in *25th Australasian Conference on Information Systems (ACIS)*, Auckland, New Zealand.

Dang-Pham, D., Pittayachawan, S., and Bruno, V. 2015. "Investigating the formation of information security climate perceptions with social network analysis: a research proposal," in *19th Pacific Asia Conference on Information Systems (PACIS)*, Singapore.

Dong, S., and Yang, X. 2009. "An improved motivation model for people behaviors change in virtual communities based on social cognitive theory," in *2009 1st International Conference on Information Science and Engineering, ICISE 2009*, pp. 2274–2277.

Dourish, P., Grinter, R. E., Delgado de la Flor, J., and Joseph, M. 2004. "Security in the wild: user strategies for managing security as an everyday, practical problem," *Personal and Ubiquitous Computing* (8:6), pp. 391–401.

Dubé, L., and Paré, G. 2003. "Rigor in Information Systems Positivist Case Research: Current Practices, Trends, and Recommendations," *MIS Quarterly* (27:4), pp. 597–635.

Gartner. 2015. "Gartner Security & Risk Management Summit 2015," National Harbor, MD (available at http://www.gartner.com/binaries/content/assets/events/keywords/security/sec22/sec21tripreport.pdf).

Gerring, J. 2004. "What Is a Case Study and What Is It Good for?," *American Political Science Review* (98:02), pp. 341–354.

Goo, J., Yim, M., and Kim, D. 2014. "A Path to Successful Management of Employee Security Compliance: An Empirical Study of Information Security Climate," *IEEE Transactions on Professional Communication* (57:4), pp. 1–24.

Guo, K. H., and Yuan, Y. 2012. "The effects of multilevel sanctions on information security violations: A mediating model," *Information & Management* (49:6), Elsevier B.V., pp. 320–326.

Hagger, M. S., and Chatzisarantis, N. L. D. 2005. "First- and higher-order models of attitudes, normative influence, and perceived behavioural control in the theory of planned behaviour," *The British journal of social psychology / the British Psychological Society* (44:Pt 4), pp. 513–535.

Herath, T., and Rao, H. R. 2009. "Protection motivation and deterrence: a framework for security policy compliance in organisations," *European Journal of Information Systems* (18:2), pp. 106–125.

Hossain, L., and de Silva, A. 2009. "Exploring user acceptance of technology using social networks," *Journal of High Technology Management Research* (20:1), Elsevier Inc., pp. 1–18.

ISO. 2015. "ISO/IEC 27001–Information security management," *Management system standards* (available at http://www.iso.org/iso/home/standards/management-standards/iso27001.htm).

Jaafar, N. I., and Ajis, A. 2013. "Organizational Climate and Individual Factors Effects on Information Security Faculty of Business and Accountancy," *International Journal of Business and Social Science* (4:10), pp. 118–130.

James, L. R., Choi, C. C., Ko, C.-H. E., McNeil, P. K., Minton, M. K., Wright, M. A., and Kim, K. 2008. "Organizational and psychological climate: A review of theory and research," *European Journal of Work and Organizational Psychology* (17:1), pp. 5–32.

Jones, A., and James, L. 1979. "Psychological climate: Dimensions and relationships of individual and aggregated work environment perceptions," *Organizational behavior and human performance* (250), pp. 201–250.

Karjalainen, M., and Siponen, M. 2011. "Toward a New Meta-Theory for Designing Information Systems (IS) Security Training Approaches," *Journal of the Association for Information Systems* (12:8), pp. 518–555.

Ten Kate, S., Haverkamp, S., and Feldberg, F. 2010. "Social network influences on technology acceptance : A matter of tie strength, centrality and density," *23rd Bled eConference eTrust*, pp. 18–32.

Laws, K., and McLeod, R. 2006. "Case study and grounded theory: Sharing some alternative qualitative research methodologies with system professionals," in *Proceedings of the 22nd International Conference of the Systems Dynamics Society*, pp. 1–25.

Liang, H., Wei, K. K., and Xue, Y. 2010. "Understanding the Influence of Team Climate on IT Use," *Journal of the Association for Information Systems* (11:8), pp. 414–432.

Neal, A., Griffin, M., and Hart, P. 2000. "The impact of organizational climate on safety climate and individual behavior," *Safety Science* (34), pp. 99–109.

Riege, A. M. 2003. "Validity and reliability tests in case study research: a literature review with 'hands-on' applications for each research phase," *Qualitative Market Research: An International Journal* (6:2), pp. 75–86.

Rosso, B. D., Dekas, K. H., and Wrzesniewski, A. 2010. "On the meaning of work: A theoretical integration and review," *Research in Organizational Behavior* (30:C), Elsevier Ltd, pp. 91–127.

Saint-Charles, J., and Mongeau, P. 2009. "Different relationships for coping with ambiguity and uncertainty in organizations," *Social Networks* (31:1), pp. 33–39.

Schneider, B., Ehrhart, M. G., and Macey, W. H. 2013. "Organizational climate and culture.," *Annual review of psychology* (64), pp. 361–88.

Schneider, B., and Reichers, A. 1983. "On the etiology of climates," *Personnel psychology* (1934), pp. 19–40.

Siponen, M., Adam Mahmood, M., and Pahnila, S. 2014. "Employees' adherence to information security policies: An exploratory field study," *Information & Management* (51:2), Elsevier B.V., pp. 217–224.

Sykes, T., Venkatesh, V., and Gosain, S. 2009. "Model of acceptance with peer support: A social network perspective to understand employees' system use," *MIS quarterly* (33:2), pp. 371–393.

Thomson, K., Solms, R. Von, and Louw, L. 2006. "Cultivating an organizational information security culture," *Computer Fraud & Security* (October), pp. 49–50.

Tichy, N. M., Tushman, M. L., and Fombrun, C. 1979. "Social Network Analysis for Organizations," *The Academy of Management Review* (4:4), pp. 507–519.

Trafimow, D., Sheeran, P., Conner, M., and Finlay, K. a. 2002. "Evidence that perceived behavioural control is a multidimensional construct: perceived control and perceived difficulty.," *The British Psychological Society* (41:1), pp. 101–121.

Yin, R. K. 2009. *Case study research design and methods*.

Zaidah, Z. 2007. "Case study as a research method," *Jurnal Kemanusiaan* (9), pp. 1–6.

Zohar, D. 2013. "Safety Climate: Conceptualization, Measurement, and Improvement," in *The Oxford Handbook of Organizational Climate and Culture*, pp. 317–334.

Zohar, D., and Tenne-Gazit, O. 2008. "Transformational leadership and group interaction as climate antecedents: a social network analysis.," *The Journal of applied psychology* (93:4), pp. 744–757.